\documentclass[aps,prb,twocolumn,amsmath,amssymb,groupedaddress,floatfix]{revtex4-1}
 \usepackage{times}
\usepackage{graphicx}
\usepackage{amsmath}
\usepackage{amstext}
\usepackage{amssymb}
\usepackage[colorlinks,citecolor=blue]{hyperref}
\usepackage{graphicx}
\usepackage{amsmath}
\usepackage{amstext}
\usepackage{amssymb}
\usepackage{amsfonts}
\usepackage{longtable,booktabs}
\usepackage{hyperref}\usepackage{url}
\usepackage{subfigure}%
\usepackage{dsfont}

\usepackage{amsbsy}
\usepackage{dcolumn}
\usepackage{amsthm}
\usepackage{bm}
\usepackage{esint}
\usepackage{multirow}
\usepackage{hyperref}
\hypersetup{
    colorlinks=true,
    linkcolor=blue,
    filecolor=magenta,
    urlcolor=cyan,
}
\usepackage{cleveref}

\usepackage{mathrsfs}
\usepackage{amsfonts}
\usepackage{amsbsy}
\usepackage{dcolumn}
\usepackage{bm}
\usepackage{multirow}
\usepackage{color}
\def\Z{\mathbb{Z}}

  \usepackage{extarrows}
 \usepackage{graphicx}
    \usepackage{amssymb}
    \usepackage{amsthm}
    \usepackage{mathtools}
    \usepackage{adjustbox}

\usepackage{datetime}

      \theoremstyle{remark}
      
\begin{document}

   \title{Three-dimensional Anomalous Twisted Gauge Theories with Global Symmetry: Implications for Quantum Spin Liquids}
 \author{Peng Ye}
  \affiliation{Department of Physics and Institute for Condensed Matter Theory, University of Illinois at Urbana-Champaign, IL 61801, USA}
 
   \begin{abstract}  
Topological spin liquids can be described by topological gauge theories with global symmetry. Due to the presence of both nontrivial bulk deconfined gauge fluxes and global symmetry, topological spin liquids are examples of the so-called ``Symmetry Enriched Topological phases'' (SET).   In this paper,  we find that, in some twisted versions of topological gauge theories (with discrete Abelian gauge group $G_g$),  implementing a global symmetry (denoted by $G_s$) is  anomalous although symmetry charge carried by topological point-like excitations is normally fractionalized and classified by the second cohomology group. To demonstrate the anomaly, we fully gauge the global symmetry, rendering a new gauge theory that is not gauge invariant.  Therefore, the SET order of the ground state is anomalous, which cannot exist in 3D system alone. Such anomalous state construction generalizes the ``2D surface topological order'' to 3D. A concrete example with $G_g=\Z_2\times\Z_4$ and $G_s=\Z_2$ is calculated.   \end{abstract}
  \maketitle

Electron spins in quantum spin liquids (QSL) point in many different directions simultaneously \cite{balents_review}. It results in absence of any conventional ordered patterns (e.g., spin density waves). Despite featureless patterns of orders, \textit{topological} QSLs  in two dimensions may host emergent excitations---anyons. Braiding them mutually leads to a set of braiding statistics data. However, practically, it is very challenging to perform such braiding experiments. For QSLs that respect a certain symmetry, there are conventional experiments to diagnose them since the quantum numbers carried by anyons may couple to injected objects (e.g., neutrons in neutron scattering experiments). As a result, it is of theoretical interest to explore how symmetry enriches QSLs, which may help us design and guide experiments on characterizing two-dimensional QSLs.  This line of thinking motivated the theoretical development of ``symmetry-enriched topological phases'' (SET) \cite{Wen2002}.

SETs are long-range entangled quantum matters where  bulk fractionalized excitations (due to the existence of topological order)  may carry fractionalized quantum number of some global symmetry.  In two dimensions (2D), the mathematical framework of SET phases has been established \cite{Barkeshli14arxiv} (see, e.g., a recent review \cite{chen_review}).    However, discussions of 3D SET phases are rare.  On the experimental side, there are several  realistic proposals of $\mathbb{Z}_2$ spin liquids, such as the so-called Kitaev spin liquid state in the lattices of $\beta$- and $\gamma$-$\mathrm{Li_2IrO_3}$ type \cite{kitaev_3d,kitaev_kim,kitaev_3d_1,exp,exp_0,exp_1,exp_2}. If an unbroken spin symmetry is considered,  the   ground state should exhibit  an  {SET} order.    Theoretically, some attempts  have been made, such as Ref.~\cite{3dset_wang,3dset_xu,spt10,3dset_fidkowski,3dset_cheng,3dset_chen,3dset_ye,3dset_ye1,spt11}. 
 
Description of an SET phase requires the knowledge of bulk topological order. Although a full knowledge of 3D topological orders is lacking, there is a subset that can be studied analytically. In the subset, all topological orders are described by twisted   gauge theories \cite{spt15} of a discrete gauge group $G_g=\Z_{N_1}\times\Z_{N_2}\times\cdots$. In the field-theoretic expression,  the action is given by:
 \begin{align}
S=&i\sum_I \frac{N_I}{2\pi} \int b^I\wedge d a^I+i\sum_{IJK}\frac{q^{IJK}}{4\pi^2} \int  a^I\wedge a^J\wedge d a^K\nonumber\\
&+i\sum_{IJKL}\frac{ {t}^{IJKL}}{8\pi^3} \int a^I\wedge a^J\wedge a^K\wedge a^L\,,\label{eq:action_of_pure_gauge}
 \end{align}
where $\{q^{IJK}\}$ and $\{ t^{IJKL}\}$ are two sets of coefficients which are quantized and compactified. $\{b^I\}$ and $\{a^I\}$ are a set of 2-form and 1-form gauge fields, respectively.     Recently a lot of progress has been made based on these topological terms in gauge theories   as well as SPTs (symmetry-protected topological phases) \cite{spt6,wang_gu_wen_15,spt15,aada_ryu1,aada_ryu2,aada_wang,3dset_twisted,3dset_twisted1,ye16e,3dset_ye1,ye17b,add1,add2,add3,add4}.  All gauge theories are uniquely labeled by the coefficients $\{q, t\}$ and one-to-one correspond to Dijkgraaf-Witten lattice model \cite{dw1990} and cohomology group: 
$  {H}^{4}(G_g,\mathrm{U(1)})= \prod_{I<J}(\mathbb{Z}_{N_{IJ}})^2\!\times\!\prod_{I<J<K}(\mathbb{Z}_{N_{IJK}})^2 \times\prod_{I<J<K<L}\mathbb{Z}_{N_{IJKL}}\,,
  $where $N_{IJ,...}$ is the greatest common divisor of $N_I, N_J,\cdots$.   When all $q$'s and $t$'s are turned off, the theory reduces to the usual (i.e., untwisted) gauge theory that is described by the BF term ($\sim b^I\wedge da^I$) only.

   Recently, Ref.~\cite{3dset_twisted}  provided a potentially feasible and systematic approach to classification and characterization of 3D SETs whose topological orders are described by twisted gauge theories (\ref{eq:action_of_pure_gauge}). On-site unitary Abelian symmetry group $G_s=\Z_{K_1}\times\Z_{K_2}\times\cdots$ or $G_s=\mathrm{U(1)}\times \Z_{K_1}\times\cdots$ were considered \cite{3dset_twisted}. Later, a systematic classification of SETs was obtained \cite{3dset_twisted1}.  There is an interesting phenomenon. In some cases, only the untwisted gauge theory can have SET orders after symmetry is imposed, while twisted ones do not have SET orders. The underlying mechanism and the physical explanation are still unknown. In this paper, we aim to study this problem in details and prove that some twisted gauge theories may be incompatible with a given global symmetry $G_g$ due to the presence of an anomaly. Thus, the underlying topological QSLs are not realizable.  In order to show the anomaly, we fully gauge the global symmetry and obtain a new gauge theory. In latter, we find that gauge invariance is manifestly violated, leading to gauge anomaly. We explain the anomaly through a concrete example: $G_g=\Z_2\times \Z_4$ and $G_s=\Z_2$.

 We start with the following action of a twisted gauge theory with gauge group $G_g=\Z_2\times \Z_4$:
 \begin{align}
S=&\int i\frac{2}{2\pi}b^1\wedge da^1+\int i \frac{4}{2\pi}b^2\wedge da^2\nonumber\\
&+\int i\frac{q}{4\pi^2}a^1\wedge a^2\wedge da^2\,,\label{main_action}
\end{align}
where the first and second terms are the usual BF terms that determine the gauge group $G_g$. The last term is the twisted term that couples two discrete gauge theories together. The coefficient $q$ is not arbitrary. Instead, it is quantized and periodically identified:
\begin{align}
q=0\text{ mod } 8 \text{ or   } q=4 \text{  mod  }8\,.
\end{align}
Therefore, for the twisted term incorporated in Eq.~(\ref{main_action}), there are two choices of topologically distinct coefficients, which corresponds to two different twisted gauge theories with $G_g=\Z_2\times\Z_4$. In fact, there are in total $(N_{12})^2$ distinct gauge theories where $N_{12}$ is the greatest common divisor of $N_1$ and $N_2$. They are labeled uniquely by a pair of integers $(q,\bar q)$.  Here, $\bar q$ is the coefficient of the twisted term $a^2a^1da^1$ that is not included in the action (\ref{main_action}). In other words, the action (\ref{main_action}) corresponds to a set of twisted gauge theories labeled by $(q,0)$. The quantization and periodicity of the two integers are given by:
\begin{align}
&q=k\frac{N_1N_2}{N_{12}}\,, k\in\Z_{N_{12}}\,,\label{quantized_1}\\
&\bar q=k'\frac{N_1N_2}{N_{12}}\,, k'\in\Z_{N_{12}}\,.\label{quantized_2}
\end{align}
   In the following, let us consider non-zero $q=4\text{ mod }8$ ($N_1=2,N_2=4$) and vanishing $\bar q=0$.

The general derivation of Eqs.~(\ref{quantized_1},\ref{quantized_2}) can be found in \cite{3dset_twisted}  The key observation is that the gauge transformations of this twisted gauge theory are defined in an unusual way: 
\begin{align}
&a^I\longrightarrow a^I +d\chi^I\,,\,\\
&b^I\longrightarrow b^I+dV^I-\frac{ q}{2\pi N^I} \epsilon^{IJ3} \chi^J   da^2\,,
\end{align}
where $\epsilon^{123}=-\epsilon^{213}=1$.   
It is clear that   the usual gauge transformations of  $b^I$  are modified through adding a $q$-dependent term. As usual,  the gauge parameters $\chi^I$ and $V^I$ satisfy the following conditions:
\begin{align}
\frac{1}{2\pi}\int_{\mathcal{M}^1} d\chi^I \in\Z, \quad \frac{1}{2\pi}\int_{\mathcal{M}^2} dV^I \in\Z \,.
\end{align}
   By requiring that the Dirac quantization conditions of $b^I$ are unbroken, i.e., 
  \begin{align}
\frac{1}{2\pi}  \int_{\mathcal{M}^3}db^I\in\Z \,, \end{align}
  the coefficients $q,\bar q$ should be properly quantized. On the other hand,   the periodicity is due to a hidden shift symmetry that compactifies the domains.

 To assign symmetry, e.g., $G_s=\Z_2$, we add the following coupling term as an example:
\begin{align}
S_{coupling}=\int i\frac{1}{2\pi}b^2\wedge dA \label{coupling_symmetry}
\end{align}
where $A$ is the external (background) gauge field that is subject to the following constraint:
\begin{align}
\int_L A=0,\pm \pi, \pm 2\pi, \cdots,\label{constraint_z2_flux}
\end{align}
for any spacetime loops $L$.
In $S_{coupling}$, $A$ minimally couples to the topological current:
\begin{align}
J=\frac{1}{2\pi} \star db^2
\end{align}
\emph{Physically, this 1-form current represents the particle current in the $\Z_4$ gauge theory [see the second term in Eq.~(\ref{main_action})]. The coupling term (\ref{coupling_symmetry}) means that all $\Z_4$ gauge charge excitations carry $\Z_2$ symmetry charge while $\Z_2$ gauge charge excitations are not charged under $\Z_2$ symmetry}.  To be much clearer, we may introduce quasiparticle current $j$ of the $\Z_4$ gauge group, which minimally couples to $a^2$:
\begin{align}
S_{excitation}=\int  i j\wedge \star a^2+\int  i \Sigma \wedge \star b^2+\cdots\,,\label{coupling_excitation}
\end{align}
where $\Sigma$ is 2-form current variable for loop excitations in the $\Z_4$ gauge theory. $\cdots$ denotes excitations in the $\Z_2$ gauge theory. Note that all omitted excitations do not couple to $a^2$. We may further  integrate  over $b^2$ in the action $S+S_{coupling}+S_{excitation}$.  
Then, $a^2$ can be formally resolved by $a^2=-\frac{\pi}{2}\frac{*d}{\hat{\Delta} }\Sigma-\frac{1}{4}A\,$, where the Laplacian operator $\hat{\Delta}\equiv *d*d$.  Plugging this expression into the first term of Eq.~(\ref{coupling_excitation}), we obtain the following effective action about excitations in the presence of symmetry twist: 
\begin{align}
-i\frac{1}{4}\int A\wedge \star j+i \frac{2\pi}{4} \int j\wedge d^{-1} \Sigma\,.
\end{align}
 In this effective action, the second term    characterizes the   $\Z_4$ topological order with charge-loop braiding phase $e^{i\frac{\pi}{2}}$. Mathematically, this is a Hopf term and represents the long-range Aharonov-Bohm statistical interaction between gauge fluxes  and particles.  $d^{-1}:=\frac{d}{\hat{\Delta} }$. The first term of this effective action indicates that the unit $\Z_4$ gauge charge excitation carries $1/4$ symmetry charge of the symmetry group $\Z_2$. However, we must be more careful to achieve the conclusion of symmetry-fractionalization.

 It is generically possible that a fractional charge may be indistinguishable from an integer charge. Mathematically, the symmetry-fractionalization is classified by the second cohomology group: $H^2(\Z_2,\Z_4)=\Z_2$, which implies that there are two sets of topologically distinct patterns of symmetry fractionalization on $\Z_4$ gauge charge excitations:
\begin{align} 
\cdots\sim-\frac{1}{4}\sim& \frac{1}{4}\sim \frac{3}{4}\sim\cdots \nonumber\\
\cdots\sim -1\sim -\frac{1}{2}\sim & 0 \sim \frac{1}{2}\sim 1\sim\cdots \nonumber
\end{align}
Therefore, in the present case, half-charge is indistinguishable from integer charge. Fortunately, $1/4$ charge is still distinguishable from integer charge. Physically, this phenomenon can be simply understood via the thought experiment in which $\Z_2$ symmetry flux is inserted and a unit $\Z_4$ gauge charge excitation moves around the symmetry flux. Due to the possible attachment of gauge flux \emph{onto} symmetry flux, the experimental data (i.e., Aharonov-Bohm phase) have ambiguity that leads to the above two set of patterns of fractionalized charge.

 It seems that there is no obvious anomaly in the patterns of symmetry-fractionalization.  So far so good. In order to examine whether or not a global symmetry is imposed in an anomaly-free way,  we    fully gauge the global symmetry. If the resulting new gauge theory is well-defined (e.g., at least gauge invariant), the symmetry implementation is anomaly-free. Otherwise, symmetry implementation is anomalous and the resulting new gauge theory admits gauge anomaly. In the following, we present the details of the gauging process. The action is given by:
 \begin{align}
S=&\int i\frac{2}{2\pi}b^1\wedge da^1+\int i \frac{4}{2\pi}b^2\wedge da^2\nonumber\\
&+\int i\frac{q}{4\pi^2}a^1\wedge a^2\wedge da^2+\int i\frac{1}{2\pi}b^2\wedge dA\nonumber\\
&+\int i\frac{2}{2\pi}  B\wedge dA\,,\label{sss1122}
\end{align}
where gauge field $A$ is now considered as a dynamical gauge field rather than background gauge field.  $B$ is another dynamical 2-form gauge field that enforces the $\Z_2$ gauge fluxes of $A$ as shown in Eq.~(\ref{constraint_z2_flux}). The action can be rewritten as the following form:
 \begin{align}
S=&   \int i\frac{1}{2\pi}\left(\begin{matrix} B& b^2&b^1\end{matrix}\right) \left(\begin{matrix}
2&0&0\\
1&4&0\\
0&0&2\end{matrix}\right)\left(\begin{matrix} A\\ a^2\\a^1\end{matrix}\right)
\nonumber\\
&+\int i\frac{q}{4\pi^2}a^1\wedge a^2\wedge da^2 \label{matrix_twist_action}
\end{align}
with $q=4\text{  mod  } 8$. 
Since now all gauge fields in the action are fully dynamical, one can apply general linear transformations $GL(3,\Z)\times GL(3,\Z)$ on two-form and one-form gauge fields independently in order to   send the above theory to its canonical form:
\begin{align}
& W=\left(\begin{matrix}
1&-1&0\\
-1&2&0\\
0&0&1\end{matrix}\right)\,,\,\,\,\\
&\Omega=\left(\begin{matrix}
1&0&0\\
4&1&0\\
0&0&1\end{matrix}\right)\,,  \\
&W \left(\begin{matrix}
2&0&0\\
1&4&0\\
0&0&2\end{matrix}\right)  \Omega^T= \left(\begin{matrix}
1&0&0\\
0&8&0\\
0&0&2\end{matrix}\right) \,.
 \end{align}
 In the new basis, we have three 2-form gauge fields: $B^1, B^2, B^3$ and three 1-form gauge fields: $A^1,A^2,A^3$. They are related to the original variables ($b^1,b^2,B$ and $a^1,a^2, A$) via:
 \begin{align}
 &\left(\begin{matrix} B\\ b^2\\b^1\end{matrix}\right)=W^T\left(\begin{matrix} B^3\\ B^2\\B^1\end{matrix}\right)\,,\\
 &\left(\begin{matrix} A\\ a^2\\a^1\end{matrix}\right)=\Omega^T\left(\begin{matrix} A^3\\ A^2\\A^1\end{matrix}\right)
 \end{align}
 As a result, the twisted term in Eq.~(\ref{matrix_twist_action}) is transformed to:
 \begin{align}
 S_{twist}=\int i \frac{q}{4\pi^2} A^2 \wedge A^1\wedge dA^1
 \end{align}
 Together with the BF term in the canonical form, we obtain the total action in the new basis:
 \begin{align}
 S=&\int i\frac{8}{2\pi}B^2\wedge dA^2+\int i\frac{2}{2\pi}B^1\wedge dA^1\nonumber\\
 &+\int i \frac{q}{4\pi^2} A^2 \wedge A^1\wedge dA^1\label{gauged_action}
 \end{align}
 where we have omitted the trivial term: $i\int \frac{1}{2\pi}B^3\wedge dA^3$. Therefore, the resulting new gauge theory is a $\Z_2\times \Z_8$ gauge theory with a twisted term. According to Eqs.~(\ref{quantized_1},\ref{quantized_2}), the coefficient $q$ should be quantized as: either $q=8\text{  mod  }16$ or $q=0\text{  mod  }16$ such that the new gauge theory is gauge invariant. However, the initial value of $q$ before gauging is $q=4\text{ mod } 8$ that fits neither $8\text{ mod }16$ nor $0\text{ mod }16$. In other words, one cannot find integers $k,k',k''$ such that either   $4+8k=16k'$ or $4+8k=8+16k''$ holds. Therefore,  after gauging, we find that gauge invariance is manifestly broken in the new gauge theory, indicating a gauge anomaly. 
  
  We conclude that:
 \begin{itemize}
\item The SET phase (i.e., topological QSL) described by the action (\ref{sss1122}) is anomalous. It cannot exist alone in 3+1D.
\item The new gauge theory described by the action (\ref{gauged_action}) has gauge anomaly. It cannot exist alone in 3+1D. 
\end{itemize}

 Recalling that in 2+1D Abelian Chern-Simons theory on a spin manifold the coefficient (i.e. level) is quantized at integer $k$, represented by the notation $U(1)_k$. However, on the surface of a 3D \emph{gauged} topological insulator, the Chern-Simons term of the background gauge field has an anomalous half-level. In the present case, we may denote the twisted gauge theory (\ref{gauged_action}) by $(\Z_2\times\Z_8)_{q}$ where $q$ takes value $4\text{  mod }8$ that is half of the normal one $8\text{  mod  }16$. 
 In Ref.~\cite{kap_anomaly,cho_teo_ryu}, 2D anomalous SETs are studied.  Espeically, in Ref.~\cite{cho_teo_ryu},    2D anomalous SETs (i.e., ``surface topological order'') with $G_g=\Z_2$ and $G_s=G_1\times G_2$ are considered, where gauging $G_1$ necessarily breaks $G_2$. Our results demonstrate  anomaly in  SETs in 3D with $G_g=\Z_2\times \Z_4$ and $G_s=\Z_2$.  In Ref.~\cite{cho_teo_ryu}, such 2D anomalous SETs are conjectured as a boundary of  3D SPTs. Here, we conjecture  that:
 \begin{itemize}
\item The anomalous SET phase described by the action (\ref{sss1122})  may appear as a boundary state of a (4+1)D bulk SPT phase.
\item The  anomalous gauge theory described by the action (\ref{gauged_action})   may appear as a boundary state of a (4+1)D gauge theory (topological order state). 
\end{itemize}

It will be interesting to construct such higher dimensional topological quantum field theories, which is left to future work. It will also be interesting to study the anomaly by the Dijkgraaf-Witten lattice model realization of the action (\ref{sss1122}).  
 We expect the findings on anomaly will further shed lights on constraints on low-energy theory of topological QSLs in three dimensions.

\textit{Acknowledgement.---} 
I am grateful to  S.-Q. Ning, Z.-X. Liu, and Y. M. Lu for their  discussions during the preparation.  Part of this work was done in Banff and KITP. This work was supported in part by the NSF through grant DMR 1408713 and DMR 1725401 at the University of Illinois and and grant of the Gordon and Betty Moore Foundation EPiQS Initiative through Grant No. GBMF4305.

\end{document}